\documentclass[twocolumn, twocolappendix]{aastex6.3.1}

\usepackage{amsmath}
\usepackage{amssymb}
\usepackage{physics}
\usepackage{float}
\usepackage[caption=false]{subfig}
\usepackage{booktabs}
\usepackage{hyperref}
\usepackage{verbatim}
\usepackage{natbib}

\usepackage{tabularx}

\usepackage[]{footmisc}

\usepackage{color}

\newcommand{\pvalue}{$p$-value}

\begin{document}

\title{Forecasting pulsar timing array sensitivity to anisotropy in the stochastic gravitational wave background}

\author{Nihan Pol}
\author{Stephen~R.~Taylor}
\affil{Department of Physics and Astronomy, Vanderbilt University, 2301 Vanderbilt Place, Nashville, TN 37235, USA}
\author{Joseph~D.~Romano}
\affil{Department of Physics and Astronomy, Texas Tech, etc}

\begin{abstract}
    Statistical anisotropy in the nanohertz-frequency gravitational-wave background (GWB) is expected to be detected by pulsar timing arrays (PTAs) in the near future. By developing a frequentist statistical framework that intrinsically restricts the GWB power to be positive, we establish scaling relations for multipole-dependent anisotropy decision thresholds that are a function of the noise properties, timing baselines, and cadences of the pulsars in a PTA. We verify that $(i)$ a larger number of pulsars, and $(ii)$ factors that lead to lower uncertainty on the cross-correlation measurements between pulsars, lead to a higher overall GWB signal-to-noise ratio, and lower anisotropy decision thresholds with which to reject the null hypothesis of isotropy. Using conservative simulations of realistic NANOGrav datasets, we predict that an anisotropic GWB with angular power $C_{l=1} > 0.3\,C_{l=0}$ may be sufficient to produce tension with isotropy at the $p = 3\times10^{-3}$ ($\sim3\sigma$) level in near-future NANOGrav data with a $20$~yr baseline. We present ready-to-use scaling relationships that can map these thresholds to any number of pulsars, configuration of pulsar noise properties, and sky coverage. We discuss how PTAs can improve the detection prospects for anisotropy, as well as how our methods can be adapted for more versatile searches.
    
\end{abstract}

          
\section{Introduction}
    
    All of the long-baseline pulsar timing arrays (PTAs)---namely the North American Nanohertz Observatory for Gravitational Waves \citep[NANOGrav,][]{NANOGrav}, the European Pulsar Timing Array \citep[EPTA,][]{EPTA} and the Parkes Pulsar Timing Array \citep[PPTA,][]{PPTA}---have now reported evidence for the presence of a spectrally-common process in their latest datasets \citep{NG12p5_detection, epta_dr2_gwb, ppta_dr2_gwb}. However, the evidence for Hellings \& Downs (HD) cross correlations \citep{HD}, which is considered the definitive signature of a stochastic gravitational wave background (GWB), was not significant in any of these datasets. A spectrally-common process was also reported by the International Pulsar Timing Array \citep[IPTA,][]{IPTA} consortium in their second data release \citep[DR2,][]{ipta_dr2_dataset}, which used older versions of datasets from the aforementioned PTAs. However this dataset also lacked definitive evidence for the HD signature \citep{ipta_dr2_gwb}. Despite this, the detection of such a spectrally-common process is considered as the first step towards the eventual detection of a GWB \citep{romano_ac_v_cc}. Based on simulations of the NANOGrav 12.5 yr dataset \citep{ng12p5_timing}, and if the signal observed in the 12.5 yr dataset is an astrophysical GWB signal, NANOGrav is expected to have sufficient evidence to report the detection of HD-consistent correlations within the next few years \citep{astro4cast}. 
    
    Once this signal is confirmed to be a GWB signal in future PTA datasets, the onus will be on characterizing the source of the GWB. An important part of this analysis will be measuring the spatial distribution of power in the GWB. For example, if the source of the GWB is a population of inspiraling supermassive black hole binaries \citep[SMBHBs,][]{Rajagopal1995, Jaffe2003, Sesana2004, BurkeSpolaor2019}, then their spatial distribution might track that of the local matter distribution. In particular, nearby galaxy clusters that host an overabundance of SMBHBs may show up as a hotspot of GW emission on the sky. For example, the Virgo cluster \citep{virgo} has an angular diameter of $\sim$10$^{\circ}$, and could show up as a hotspot on the GW sky at a multipole of $l_{\rm Virgo} = 180^{\circ} / \theta \approx 18$. On the other hand, a single SMBHB that is louder than the GWB will show up as a point source anisotropy, with multiple such single sources producing a ``popcorn''- or pixel-style spatial distribution of GWB power. However, if the GWB is produced by a cosmological source, such as cosmic strings \citep[e.g.,][]{cosmic_string_spectrum} or primoridal GWs \citep[e.g.,][]{primordial_gw_spectrum}, then the GWB power distribution on the sky may not display the same anisotropies as that from a SMBHB-produced GWB.
    Thus, the anisotropy of the GWB, or the lack of it, allows us to make inferences about the source of the GWB.
    
    Multiple techniques have been developed to probe the anisotropy of a GWB with PTAs \citep[e.g.,][]{cornish_eigenmaps, ming_pta_anis, taylor_pta_anis, bumpy_bkgrnd}. While these methods differ in their choice of basis for modeling anisotropy, they all 
    take the pulsar times-of-arrival (TOAs) as their initial data, and employ Bayesian techniques to constrain parameters that describe anisotropy-induced deviations away from the HD signature. Constraints on anisotropy were first presented by the EPTA as part of the analysis of their first data release \citep{epta_anisotropy}, where they showed that the strain amplitude in $l > 0$ spherical-harmonic multipoles is less than $40\%$ of the monopole value ($l = 0$, i.e., isotropy).
    
    
    As PTA datasets grow longer in timespan, add more pulsars, and become denser with higher cadence observations, the analysis time for Bayesian methods based on TOAs are going to increase dramatically. Additionally, Bayesian model selection of anisotropy is predicated on having an appropriate hypothesis of the anisotropy, e.g., a power map built from galaxy catalogs, statistically populated with inspiraling SMBHBs. In other words, Bayesian model selection always requires two models to compare; a null and signal model. By contrast, frequentist techniques allow one to reject a null hypothesis (in this case, isotropy) if the data are in sufficient tension with it. Importantly, rejecting a null hypothesis at a probability given by some $p$-value is not the same as accepting a signal hypothesis with a certainty of $1 - p$. However significant tension with the assumption of isotropy is an important indicator of beyond-HD signatures in the cross-correlation data.
    
    To overcome these challenges, we develop a frequentist framework that employs the cross-correlations between pulsar timing residuals across a PTA as sufficient data with which to search for anisotropy. These cross correlations are measured as part of the standard GWB detection pipelines \citep[e.g.,][]{NG12p5_detection, epta_dr2_gwb, ppta_dr2_gwb}, using established optimal two-point correlation techniques \citep{allen_romano_OS, anholm_OS, NG5yr_OS, noise_marg_os_vigeland}. Thus our framework can be easily incorporated into ongoing analysis campaigns. The data volume in our framework is also significantly lower than analyses starting at the TOA-data level, enabling rapid estimation of anisotropy. Finally, as mentioned above, the frequentist framework allows us to infer (although not necessarily claim detection of) anisotropy via rejection of the null hypothesis of isotropy, thereby simplifying the process of constraining anisotropy.
    
    Other frequentist techniques for searching for anisotropy with PTAs have been proposed. These include, for example, the Fisher matrix formalism developed in \citet{ali-hamoud_1} and \citet{ali-hamoud-2} where they employ ``principal maps'', which are the eigenmaps of the Fisher matrix, to search for anisotropy. \citet{hotinli}, on the other hand, decompose the timing residual power spectrum onto bipolar spherical harmonics to search for anisotropy using the correlations between pulsars in a PTA. These frequentist techniques also focus on the detection of anisotropy through rejection of the null hypothesis of isotropy.
    The use of cross correlations between detector baselines is also commonly used when searching for anisotropic GWBs in the LIGO \citep[e.g.,][]{thrane_ligo_anis, ligo_O3_anis_search} and LISA bands \citep[e.g.,][]{banagiri_blip}.
    
    We use our framework on simulations of idealized-PTA and near-future NANOGrav data, presenting, for the first time, projections on the sensitivity of NANOGrav and other PTAs to GWB anisotropy in the mid-to-late 2020s. As shown in \citet{astro4cast}, if the common-spectrum signal observed in the NANOGrav 12.5 yr dataset is a GWB, then NANOGrav can be expected to have sufficient statistical evidence to claim a detection of HD correlations within the next few years. By employing the same assumptions and simulation pipeline as \citet{astro4cast} to generate our simulations, we forecast the level of anisotropy that would result in a statistically significant rejection of the null hypothesis of isotropy by the NANOGrav data. 
    
    The paper is organised as follows: in Section~\ref{sec:methods}, we describe the measurement of cross-correlations and their uncertainties, along with the maximum-likelihood framework and detection statistics that are used to constrain anisotropy, while Section~\ref{sec:connection} connects the cross correlation uncertainty to the noise properties, cadence, and timing baseline of the PTA. Section~\ref{sec:sim_ideal_pta} presents scaling relations for anistoropy decision thresholds of an ``ideal PTA'' as a function of cross-correlation uncertainty and number of pulsars in a PTA, while Section~\ref{sec:sim_real_pta} presents them for a realistic PTA that is generated using the NANOGrav 12.5 yr dataset. Finally, we present a discussion of results and prospects for the future in Section~\ref{sec:discuss}. In Appendix \ref{appendix:real_os} and \ref{appendix:unc_scaling} we present derivations for computing the cross correlations in real PTA datasets and scaling relations for the cross correlation uncertainty with respect to PTA specifications like timing baseline, white noise and observation cadence.
    
\section{Methods} \label{sec:methods}
    
    \subsection{The optimal cross correlation statistic and overlap reduction function} \label{subsec:os_orf}
        
        A GWB can be uniquely identified through the cross correlations it induces between the times-of-arrivals (TOAs) of pulsars in a PTA \citep{HD, tiburzi_cc}. The timing cross correlations between pulsar pairs, $\rho_{ab}$, and their uncertainties, $\sigma_{ab}$, can be written as \citep{NG5yr_OS, optimal_statistic_chamberlin, siemens_scaling_laws, noise_marg_os_vigeland}
        \begin{align} \label{eq:cross_corr}
            \displaystyle \rho_{ab} &= \frac{\delta\textbf{t}^T_{a} \textbf{P}^{-1}_{a} \hat{\textbf{S}}_{ab} \textbf{P}^{-1}_{b} \delta\textbf{t}^T_{b}}{\textrm{tr}\left[ \textbf{P}^{-1}_{a} \hat{\textbf{S}}_{ab} \textbf{P}^{-1}_{b} \hat{\textbf{S}}_{ba} \right]}, \nonumber\\
            \displaystyle \sigma_{ab} &= \left( \textrm{tr}\left[ \textbf{P}^{-1}_{a} \hat{\textbf{S}}_{ab} \textbf{P}^{-1}_{b} \hat{\textbf{S}}_{ba} \right] \right)^{-1/2},
        \end{align}
        where $\delta\mathbf{t}_a$ is a vector of timing residuals for pulsar-$a$, $\textbf{P}_a = \langle \delta \mathbf{t}_a \delta \mathbf{t}_a^T\rangle$ is the measured autocovariance matrix of pulsar-$a$, and $\hat{\textbf{S}}_{ab}$ is the template scaled-covariance matrix between pulsar-$a$ and pulsar-$b$. This scaled-covariance matrix is a template for the GWB spectral shape only, and is independent of the GWB's amplitude and the cross-correlation signature that it induces. It is related to the full covariance matrix by $\textbf{S}_{ab} = \langle \delta \mathbf{t}_a \delta \mathbf{t}_b^T\rangle = A_{\rm gwb}^2\chi_{ab}\hat{\textbf{S}}_{ab}$, where $A_{\rm gwb}$ is the GWB amplitude corresponding to a given strain-spectrum template, and $\chi_{ab}$ is the GWB-induced cross-correlation value for this pair of pulsars, e.g., the Hellings \& Downs factor in the case of an isotropic GWB.
        
        The cross-correlation statistic accounts for fitting and marginalization over the timing ephemeris of each pulsar, as well as its intrinsic white and red noise characteristics, where a power-law spectrum is usually assumed for the intrinsic red noise. Implementations of the statistic also usually assume a power-law strain spectrum template for the GWB following $f^{-2/3}$ \citep{phinney} filtering it across the pairwise-correlated timing residuals in order to extract an optimal measurement of the GWB amplitude \citep{optimal_statistic_chamberlin,noise_marg_os_vigeland}. However, we note that this statistic is flexible enough to allow for different parametrized spectral templates. See \autoref{appendix:real_os} for details of how this is implemented in real PTA searches for the GWB, including determining the noise weighting for each pulsar, setting the GWB's spectral template, and how the parameters of the pulsar timing ephemeris are marginalized over. In this paper we simulate and analyze data at the level of $\{\rho_{ab}, \sigma_{ab} \}$, then connect our results later to the underlying geometry and noise characteristics of the PTA (see \autoref{appendix:unc_scaling}).
        
        A PTA with $N_{\rm psr}$ pulsars has $N_{\rm cc} = N_{\rm psr} (N_{\rm psr} - 1) / 2$ distinct cross-correlation values. The angular dependence of these empirically measured cross-correlations can be modelled by the detector overlap reduction function (ORF)\footnote{We note that the term ORF in ground- and space-based literature usually involves a frequency dependence. However this frequency-dependence factors out in the PTA regime, such that we use the term ORF to denote only the angular dependence of the pairwise cross-correlated data.} \citep{orf_citation, ming_pta_anis, taylor_pta_anis, Gair_pta_cmb_anis, bumpy_bkgrnd} such that, for an unpolarized GWB,
        \begin{align} \label{eq:orf}
            \Gamma_{ab} \propto \int_{S^2} d^2\hat\Omega \,\,P(\hat\Omega) &\left[ \mathcal{F}^+(\hat{p}_a,\hat\Omega)\mathcal{F}^+(\hat{p}_b,\hat\Omega) \right. \nonumber\\
            &\left. +\, \mathcal{F}^\times(\hat{p}_a,\hat\Omega)\mathcal{F}^\times(\hat{p}_b,\hat\Omega) \right],
        \end{align}
        where $P(\hat\Omega)$ is the angular power of the GWB in direction $\hat\Omega$, normalized such that $\int_{S^2}d^2\hat\Omega \,\,P(\hat\Omega) = 1$, and $\mathcal{F}^A(\hat{p},\hat\Omega)$ is the antenna response pattern of a pulsar in unit-vector direction $\hat{p}_a$ to each GW polarization $A\in[+,\times]$:
        \begin{equation}
            \displaystyle \mathcal{F}^A (\hat{p}, \hat{\Omega}) = \frac{1}{2} \frac{\hat{p}^i \hat{p}^j}{1 - \hat{\Omega} \cdot \hat{p}} e_{ij}^A (\hat{\Omega}),
            \label{eq:antenna_resp_def}
        \end{equation}
        where $e_{ij}^A(\hat{\Omega})$ are polarization basis tensors, and $(i, j)$ are spatial indices.
        
        We can recast the sky integral in \autoref{eq:orf} as a sum over equal-area pixels \citep{Gair_pta_cmb_anis, bumpy_bkgrnd}. Assuming an unpolarized GWB, and ignoring random pulsar term contributions to the cross correlations, this can be written as
        \begin{equation}
            \displaystyle \Gamma_{ab} \propto \sum_{k} P_{k} \left[\mathcal{F}^+_{a,k}\mathcal{F}^+_{b,k} + \mathcal{F}^\times_{a,k}\mathcal{F}^\times_{b,k}\right],
            \label{eq:orf_full}
        \end{equation}
        where $k$ denote pixel indices. Or, in a general matrix form
        \begin{equation}
            \displaystyle \mathbf{\Gamma} = \mathbf{R} \textbf{P},
            \label{eq:orf_matrix}
        \end{equation}
        where $\mathbf{\Gamma}$ is an $N_\mathrm{cc}$ vector of ORF values for all distinct pulsar pairs, $\mathbf{P}$ is an $N_\mathrm{pix}$ vector of GWB power values at different pixel locations, and $\mathbf{R}$ is a $(N_\mathrm{cc}\times N_\mathrm{pix})$ overlap response matrix given by
        \begin{equation}
            R_{ab,k} = \frac{3}{2N_{\rm pix}} \left[\mathcal{F}^+_{a,k}\mathcal{F}^+_{b,k} + \mathcal{F}^\times_{a,k}\mathcal{F}^\times_{b,k}\right],
        \end{equation}
        where the normalization is chosen so that the ORF matches the Hellings \& Downs values in the case of an isotropic GWB with $P_k=1\,\,\forall k$.
        
        Thus, in our notation the expected value of $\rho_{ab}$ is such that $\langle\rho_{ab}\rangle = A^2\Gamma_{ab}$. However, in the remainder of this paper we will deal with amplitude-scaled cross-correlation values, $\rho_{ab}/A^2$, where we assume that in a real search an initial fit for $A^2$ will be performed on $\{\rho_{ab}\}$ under the assumption of isotropy. These amplitude-scaled cross-correlation values can then be directly compared with the ORF model to probe anisotropy. We suppress the explicit amplitude scaling in the remainder of our notation, such that $\rho_{ab}$ henceforth implies amplitude-scaled cross-correlation values.
        
        Assuming a stationary Gaussian distribution for the cross correlation uncertainty, the likelihood function for the cross correlations can be written as
        \begin{equation}
            \displaystyle p(\boldsymbol{\rho} | \mathbf{P}) =  \frac{\textrm{exp} \left[ -\frac{1}{2} (\boldsymbol{\rho} - \mathbf{R} \mathbf{P})^T \, \mathbf{\Sigma}^{-1} \, (\boldsymbol{\rho} - \mathbf{R} \mathbf{P}) \right]}{\sqrt{\mathrm{det}(2\pi\mathbf{\Sigma})}},
            \label{eq:anis_lkl}
        \end{equation}
        where $\mathbf{\Sigma}$ is the diagonal covariance matrix of cross-correlation uncertainties. We now discuss several different bases on which to decompose the angular power vector, $\mathbf{P}$.
        
        \subsubsection{Pixel basis}
            
            The GWB power can be parametrized using HEALPix sky pixelization \citep{healpix}, where each equal-area pixel is independent of the pixels surrounding it
            \begin{equation}
                \displaystyle P(\hat{\Omega}) = \sum_{\hat{\Omega}^{\prime}} P_{\hat{\Omega}^{\prime}} \delta^2(\hat{\Omega}, \hat{\Omega}^{\prime}).
                \label{eq:pix_basis}
            \end{equation}
            The number of pixels is set by $N_{\rm pix} = 12 N_{\rm side}^2$ and $N_{\rm side}$ defines the tessellation of the healpix sky \citep{healpix}. The rule of thumb for PTAs is to have $N_{\rm pix} \leq N_{\rm cc}$ \citep{cornish_eigenmaps, romano_cornish_review}. This basis is well suited for detection of pixel-scale anisotropy, which can arise from individual sources of GWs, and where an isotropic GWB would be represented by equal power (within uncertainties) in each pixel on the sky.
            
        \subsubsection{Spherical harmonic basis}
            
            Alternatively, the GWB power can be decomposed onto the spherical harmonic basis \citep[e.g.][]{thrane_ligo_anis}, where the lowest order multipole ($l = 0$) defines an isotropic background, while higher multipoles add anisotropy. The GWB power in this basis can be written as
            \begin{equation}
                \displaystyle P(\hat{\Omega}) = \sum_{l = 0}^{\infty} \sum_{m = -l}^{l} c_{lm} Y_{lm}(\hat{\Omega}),
                \label{eq:sph_basis}
            \end{equation}
            where $Y_{lm}$ are the real valued spherical harmonics and $c_{lm}$ are the spherical harmonic coefficients of $P(\hat{\Omega})$. In this basis, the ORF anisotropy coefficient for the $l,m$ components between pulsars $a,b$ can be written as \citep{bumpy_bkgrnd},
            \begin{equation}
                \displaystyle \Gamma_{(lm)(ab)} = \kappa \sum_{k} c_{lm} Y_{lm,k} \left[ \mathcal{F}_{a,k}^{+} \mathcal{F}_{b,k}^{+} + \mathcal{F}_{a,k}^{\times} \mathcal{F}_{b,k}^{\times} \right],
                \label{eq:orf_sph_basis}
            \end{equation}
            where $k$ represents the pixel index corresponding to $\hat{\Omega}$ and the constant $\kappa$ accounts for the pixel area in the healpix sky tessellation.
            
            This basis representation, contrary to the pixel basis, is better suited for modeling large-scale anisotropies in the GWB. Based on diffraction-limit arguments, the highest order mode, $l_{\rm max}$, that can be used for modeling the anisotropy depends on the number of pulsars in the PTA, $l_{\rm max} \sim \sqrt{N_{\rm psr}}$ \citep{pen_boyle_pta_resolution, romano_cornish_review}. However, \citet{higher_lmax_limit} have shown that while the diffraction limit is attuned to maximizing the significance of the detection of anisotropy, values of $l > l_{\rm max}$ can be included in spherical harmonic decompositions to improve the localization of any anisotropy after its detection. 
            
            The results can be expressed in terms of $C_l$, which is the squared angular power in each mode $l$
            \begin{equation}
                \displaystyle C_l = \frac{1}{2l + 1} \sum_{m = -l}^{l} |c_{lm}|^2.
                \label{eq:angular_power}
            \end{equation}
            Physically, $C_l$ represents the amplitude of statistical fluctuations in the angular power of the GWB at scales corresponding to $\theta = 180^{\circ} / l$.
            An isotropic background in this basis will contain power only in the $l = 0$ multipole, thus filling the entire sky, while an anisotropic background will have power in the higher $l$ multipoles. 
            On the other hand, the variance of the angular power distribution can be written as \citep{alex_jenkins_disst}
            \begin{equation}
            \displaystyle {\rm Var}[P(\hat{\Omega})] \approx \int {\rm d (ln }\,l) \frac{l (l + 1)}{2\pi} C_l.
            \end{equation}
            The quantity $l(l + 1)C_l / 2 \pi$ thus represents the variance per logarithmic multipole bin, and is what we use to present our results in this work. As pointed out in \citet{alex_jenkins_disst}, reporting $C_l$ is analogous to reporting the GWB strain power spectral density, with $C_l = $ constant representing a white angular power spectrum, while reporting $l(l + 1)C_l / 2 \pi$ is analogous to reporting the GWB energy density spectrum $\Omega_{\rm gw}(f)$, with $l(l + 1)C_l / 2 \pi = $ constant representing a scale-invariant angular power spectrum.
            
            For these bases, since the problem is linear in the regression coefficients, the maximum likelihood solution can be derived analytically \citep{thrane_ligo_anis, romano_cornish_review, ivezic_book}
            \begin{equation}
                \displaystyle \hat{\mathbf{P}} = \mathbf{M}^{-1} \mathbf{X},
                \label{eq:max_lkl_power}
            \end{equation}
            where $\mathbf{M} \equiv \mathbf{R}^{T} \mathbf{\Sigma}^{-1} \mathbf{R}$ is the Fisher information matrix, with the uncertainties on the $c_{lm}$ coefficients given by the diagonal elements of $\mathbf{M}^{-1}$, and $\mathbf{X} \equiv \mathbf{R}^{T} \mathbf{\Sigma}^{-1} \boldsymbol{\rho}$ is the ``dirty map'', an inverse-noise weighted representation of the total power on the sky as ``seen" through the response of the pulsars in the array.
            
        \subsubsection{Square-root spherical harmonic basis} \label{subsubsec:sqrt_basis}
            
            A drawback of both the pixel and spherical harmonic bases is that they allow the GWB power to assume negative values, which is an unphysical realization of the GWB. While these tendencies can be curbed through the use of regularization techniques or rejection priors \citep{taylor_pta_anis}, this results in the addition of a hyperparameter to the analysis that requires further optimization or non-analytic priors \citep{taylor_pta_anis, bumpy_bkgrnd}. A more elegant solution is to use a basis that intrinsically conditions the GWB power to be positive over the whole sky.
            
            Such a basis can be generated by modeling the square-root of the GWB power, $P(\hat{\Omega})^{1/2}$, rather than modelling the power itself. This technique was introduced in a Bayesian context in \citet{cg_anis_ligo} for LIGO, \citet{banagiri_blip} for LISA, and in \citet{bumpy_bkgrnd} for PTAs. Decomposing the square-root power onto spherical harmonics, the GWB power can be written as
            \begin{equation}
                \displaystyle P(\hat{\Omega}) = [P(\hat{\Omega})^{1/2}]^2 = \left[ \sum_{L = 0}^{\infty} \sum_{M = -L}^{L} b_{LM} Y_{LM} \right]^2,
                \label{eq:sqrt_basis}
            \end{equation}
            where $Y_{LM}$ are real valued spherical harmonics and $b_{LM}$ are the search coefficients. \citet{banagiri_blip} showed that the search coefficients in this basis can be related to the spherical harmonic coefficients via
            \begin{equation}
                \displaystyle c_{lm} = \sum_{LM} \sum_{L^{\prime} M^{\prime}} b_{LM} b_{L^{\prime} M^{\prime}} \beta_{lm}^{LM, L^{\prime} M^{\prime}},
                \label{eq:sqrt_to_sph}
            \end{equation}
            where $\beta_{lm}^{LM, L^{\prime} M^{\prime}}$ is defined as
            \begin{equation}
                \displaystyle \beta_{lm}^{LM, L^{\prime} M^{\prime}} = \sqrt{ \frac{(2L + 1) (2L^{\prime} + 1)}{4 \pi (2l + 1)}} C^{lm}_{LM, L^{\prime} M^{\prime}} C^{l0}_{L0, L^{\prime} 0},
                \label{eq:cg_coeff}
            \end{equation}
            with $C^{lm}_{LM, L^{\prime} M^{\prime}}$ being the Clebsch-Gordon coefficients. \citet{bumpy_bkgrnd} showed that even though a full reconstruction of $c_{lm}$ requires an infinite sum over the $b_{LM}$ coefficients, restricting the maximum mode to $L \leq L_{\rm max} = l_{\rm max}$ is sufficient to produce an accurate reconstruction of the GWB power.
            
            Since the problem in this basis is non-linear in the regression coefficients, the likelihood in \autoref{eq:anis_lkl} cannot be maximized analytically. The maximum likelihood solution thus has to be calculated through numerical optimization techniques. In this work, we use the {\sc lmfit} \citep{lmfit} Python package, where we use the Levenberg-Marquardt (LM) optimization algorithm \citep{levenberg1944method, marquardt1963algorithm} to determine the maximum likelihood solution. The goodness-of-fit is assessed through the $\chi_{\rm dof}^2$ that is reported by {\sc lmfit}.
            
\begin{figure}
                \centering
                \subfloat{\includegraphics[width = 0.4\textwidth]{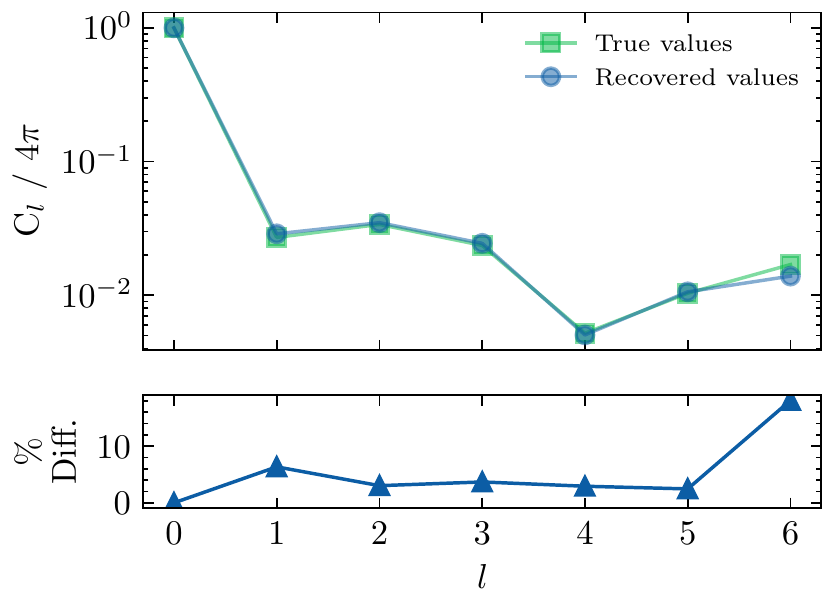}}
                \hfill
                \subfloat{\includegraphics[width = 0.4\textwidth]{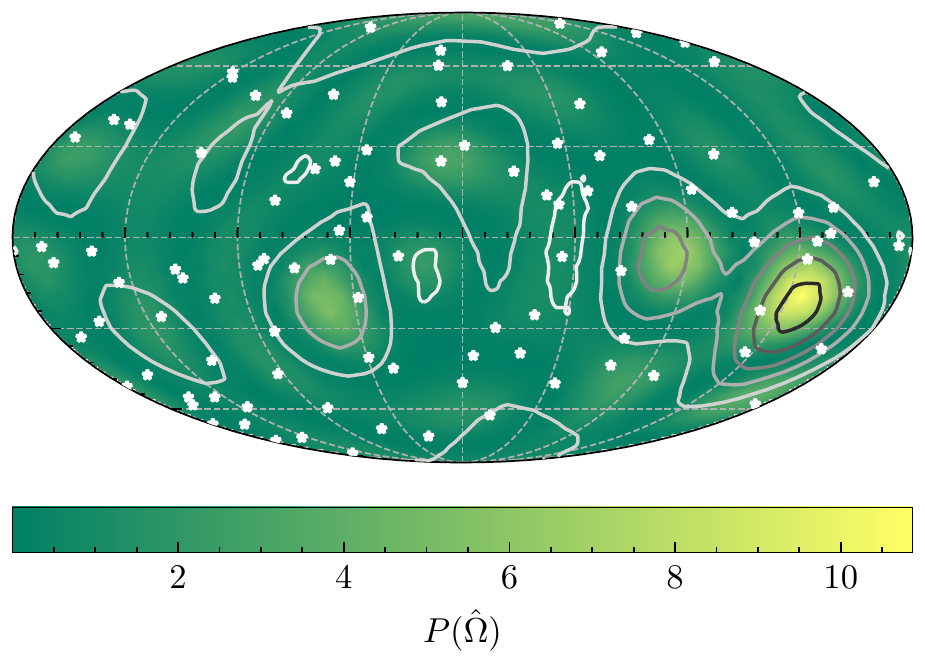}}
                \caption{Example recovery of an anisotropic GWB using the square root spherical harmonic basis described in \S\ref{sec:methods}. These simulations were based on an ideal PTA consisting of 100 pulsars, with a cross correlation uncertainty of 0.01 across all pulsar pairs, while the anisotropy is based on a realistic population of inspiraling SMBHBs \citep{bumpy_bkgrnd}. 
                 \textit{Top:} The true and recovered angular power spectrum, as well as the percent difference between them, where both are normalized such that the power in the $l = 0$ mode is $C_{0} = 4\pi$. \textit{Bottom:} The sky map of the GWB power corresponding to the recovered angular power spectrum. The contours represent the true distribution of the GWB power on the sky for the anisotropic GWB, while the stars represent the positions of the simulated pulsars on the sky.}
                \label{fig:realistic_example}
            \end{figure}

            \autoref{fig:realistic_example} shows an example of recovering an anisotropic background using this basis. To produce the simulated cross-correlation data, we inject a GW power map corresponding to the synthesized population of SMBHBs from \citet{bumpy_bkgrnd} into an ``ideal PTA'' consisting of 100 pulsars, with a constant cross-correlation measurement uncertainty of $0.01$. Note that this is an uncertainty on cross correlations that can assume any value between $-0.2$ and $0.5$, and thus represents extremely accurate measurement of the cross correlations between pulsars in the PTA (see \S\ref{subsec:define_ideal_pta} for our definition of an ``ideal PTA''). We see that this basis is capable of reproducing the injected angular power spectrum, which translates into an accurate identification of the anisotropy in the GWB. 
            
    \subsection{Detection statistics}
    
    In addition to estimating the values of the anisotropy search coefficients (i.e., pixels or spherical harmonic coefficients), we also need to quantify the evidence for the presence of anisotropy in the cross correlation data. As described earlier, when searching for anisotropy, we seek to reject the null hypothesis of isotropy. In this section, we present two frequentist detection statistics that can quantify the evidence for anisotropy by measuring the (in)compatibility of the data with the null hypothesis. 
        
        \subsubsection{Signal-to-noise ratio} \label{subsubsec:sn}
            
            Confidence in the detection of anisotropy can be quantified through the signal-to-noise (S/N) ratio, which is defined as the ratio of the maxima of the likelihood functions between any two models
            \begin{equation}
                \displaystyle \textrm{S/N} = \sqrt{2 \textrm{ln}[\Lambda_{\rm ML}]},
                \label{eq:sn_def}
            \end{equation}
            where $\Lambda_{\rm ML} = p(\boldsymbol{\rho} | \mathbf{P}_{\mathrm{ML},1})\, /\, p(\boldsymbol{\rho} | \mathbf{P}_{\mathrm{ML},2})$ is the ratio of the maxima of the likelihood functions for the two models that are under consideration.
            
            When searching for anisotropy, we can define three S/N statistics which together provide a complete description of the evidence for the GWB signal present in the cross correlation data:
            \begin{enumerate}
                \item ``Total S/N'': This is defined as the ratio of the maximum likelihood values of an anisotropic model (i.e. $p(\boldsymbol{\rho} | \mathbf{P}_{\rm ML}(l_{\rm max} > 0))$) to a model with only spatially-uncorrelated noise (i.e., $p(\boldsymbol{\rho} | \mathbf{P}_{\rm ML}=0)$). The total S/N quantifies the evidence for the presence of any signal in the cross correlations.
                \item ``Isotropic S/N'': This is defined as the ratio of the maximum likelihood values of an isotropic model (i.e., $p(\boldsymbol{\rho} | \mathbf{P}_{\rm ML}(l_{\rm max} = 0))$) to a model with noise only (i.e., $p(\boldsymbol{\rho} | \mathbf{P}_{\rm ML}~=~0)$). Note that the isotropic S/N is equivalent to the optimal S/N statistic defined in \citet{optimal_statistic_chamberlin}. The isotropic S/N ratio quantifies how well the cross correlations are described by a purely isotropic model.
                \item ``Anisotropic S/N'': This is defined as the ratio of the maximum likelihood values of a model with anisotropy (i.e., $p(\boldsymbol{\rho} | \mathbf{P}_{\rm ML}(l_{\rm max} > 0))$) to an isotropic model (i.e., $p(\boldsymbol{\rho} | \mathbf{P}_{\rm ML}(l_{\rm max} = 0))$). The anisotropic S/N ratio quantifies the evidence in favor of inclusion of modes $l > 0$.
            \end{enumerate}
        
        \subsubsection{Decision threshold}
            
            Another method for determining the significance of possible anisotropy is to assess the certainty with which the null hypothesis of isotropy can be rejected. If we can quantify the distribution of the angular power, $C_l$, under the null hypothesis, then we can also quantify how  (in)consistent the measured angular power is with isotropy through a test statistic like the \pvalue{}. 
            
            To calculate the null distribution of $C_l$ under the null hypothesis, we generate many realizations of cross-correlation data, where we assume that the measurements are Gaussian distributed around the Hellings \& Downs curve, with the spread of the distribution given by the uncertainty on the cross correlation values. We define the decision threshold, $C_l^{\rm th}$, as the value of $C_l$ corresponding to a \pvalue{} of $3\times10^{-3}$, where a measurement of angular power greater than this threshold would indicate a tension with the null hypothesis at the $\sim3\sigma$ level.
            
\section{Cross-correlation uncertainties from PTA design} \label{sec:connection}
    
    The uncertainty on the cross-correlation measurements introduced in \autoref{eq:cross_corr} depends on the pulsars that are used to construct the PTA \citep{anholm_OS, optimal_statistic_chamberlin, siemens_scaling_laws}. As shown in \citet{siemens_scaling_laws}, the trace in \autoref{eq:cross_corr} can be written as
    \begin{equation}
        \displaystyle \textrm{tr}\left[ \textbf{P}^{-1}_{a} \hat{\textbf{S}}_{ab} \textbf{P}^{-1}_{b} \hat{\textbf{S}}_{ba} \right] = \frac{2T}{A^4} \int_{f_l}^{f_h} df \frac{P_g^2(f)}{P_a(f) P_b(f)},
        \label{eq:trace_eq}
    \end{equation}
    where $T$ is the duration of time that a pulsar is observed, i.e., the timing baseline, $A$ is the amplitude of the GWB, $P_g(f)$ is the power spectrum of the timing residuals induced by the GWB, $P_a(f)$ and $P_b(f)$ are the intrinsic power spectra of pulsars $a$ and $b$, $f_l = 1/T$ and $f_h$ are the low and high frequency cutoffs used in the GWB analysis. 
    
    Using the same analysis as presented in \citet{siemens_scaling_laws} and \citet{optimal_statistic_chamberlin}, we can show that (see Appendix~\ref{appendix:unc_scaling}) in the weak-signal regime, the cross-correlation uncertainty between a given pair of pulsars scales as
    \begin{equation}
        \displaystyle \sigma \propto \frac{w^2 T^{-\gamma}}{c},
        \label{eq:weak_sig_scaling}
    \end{equation}
    while in the strong-signal regime
    \begin{equation}
        \displaystyle \sigma \propto \frac{A^2}{\sqrt{cT}},
        \label{eq:strong_sig_scaling}
    \end{equation}
    where $w$ is the white noise RMS, $c=1/\Delta t$ is the observing cadence, $\gamma = 3 - 2\alpha$ \citep{NG12p5_detection} is the slope of the timing-residual power spectrum induced by the GWB, and $A$ is the amplitude of the GWB whose characteristic strain spectrum is given by $h_c(f) = A (f / f_{\rm yr})^{\alpha}$, with $\alpha = -2/3$ for a SMBHB GWB.
    
    These scaling relations imply that the uncertainty on the cross correlations can be reduced by observing pulsars for longer duration (timing baseline), or with a higher cadence. The effect of the timing baseline and cadence is strongest in the weak-signal regime while it becomes weaker as we move into the strong-signal regime. Similarly, in the weak-signal regime, the cross-correlation uncertainty can be reduced by decreasing the intrinsic white noise by, for example, increasing the receiver bandwidth or increasing the integration time for the pulsars. The white noise does not affect the cross-correlation uncertainty in the strong-signal regime, where the GWB signal dominates over the intrinsic noise in the pulsars at all frequencies.
    
\section{Simulations with an ideal PTA} \label{sec:sim_ideal_pta}
    
    \subsection{Defining an ideal PTA} \label{subsec:define_ideal_pta}
    
        In the framework described above, the rejection of isotropy in the GWB depends on three variables: the uncertainty on the cross correlation values, the number of pulsars in the PTA (which defines the number of cross correlation values that are measured), and the distribution of the pulsars on the sky. The first two variables primarily dictate the strength of the rejection of isotropy, while the third variable is important for the characterization of the anisotropy.  
        In this section, we examine how the cross correlation uncertainty and the number of pulsars in the PTA affect an \textit{ideal} PTA, which we define as a PTA that has pulsars distributed uniformly on the sky, and all pulsars in the PTA having identical noise properties. The latter constraint implies that all measured cross-correlations have the same uncertainty. This is different from current, real PTAs, where each pulsar is unique and thus the uncertainties on the cross correlations between each pulsar pair are different. We examine realistic PTAs in \S\ref{sec:sim_real_pta}.
    
    \subsection{Scaling relations} \label{subsec:scaling_rel}
        
        For a given level of anisotropy, its detection significance will depend on the number of pulsars in the array, as well as on the accuracy with which the cross correlations between different pulsar pairs can be measured. \citet{romano_cornish_review} showed that the total S/N for linear anisotropy models can be written as 
        \begin{equation}
            \displaystyle \textrm{total S/N} = \left(\mathbf{\hat{P}}^T \left[ \mathbf{R}^T \mathbf{\Sigma}^{-1} \mathbf{R} \right]  \mathbf{\hat{P}}\right)^{1/2}, \label{eq:sn_scaling_unc}
        \end{equation}
        
        where $\mathbf{\hat{P}}$ is the maximum-likelihood estimate of the GWB power. Since $\mathbf{\Sigma}$ is a diagonal matrix of the squared cross correlation uncertainties, the total S/N $\propto \sigma^{-1}$. 
        Similarly, the total S/N is proportional to the square root of the number of data points available for inference (assuming all cross correlation uncertainties are the same for all pairs), i.e.
        \begin{equation}
            \displaystyle \textrm{total S/N} \propto \sqrt{N_{\rm cc}} = \sqrt{\frac{N_{\rm psr} (N_{\rm psr} - 1)}{2}},
            \label{eq:sn_scaling_npsr}
        \end{equation}
        where for sufficiently large $N_{\rm psr}$, the total S/N will scale linearly with the number of pulsars in the PTA.
        
\begin{figure}
            \centering
            \includegraphics[width = \columnwidth]{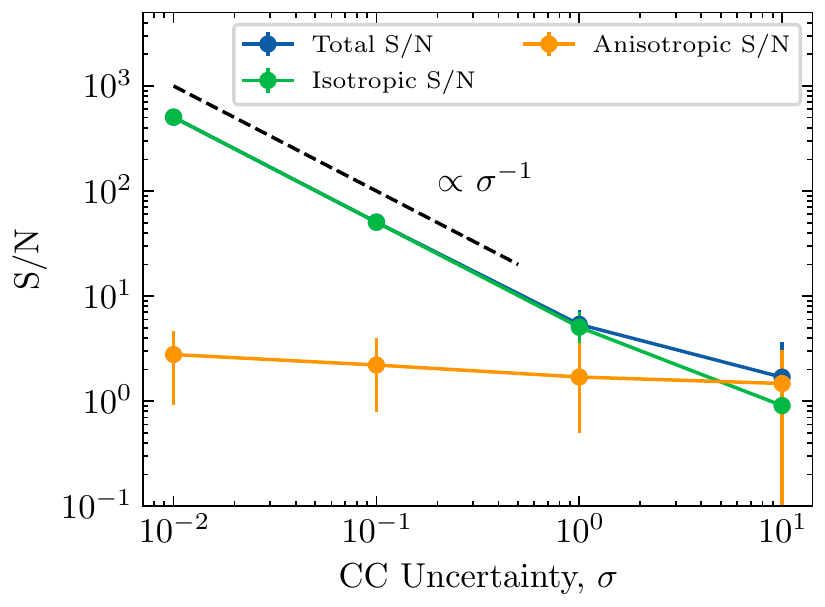}
            \caption{The evolution of the total, isotropic, and anisotropic S/N values over $10^4$ noise realizations as a function of cross-correlation uncertainty for an ideal PTA with $100$ pulsars and an isotropic GWB injection. The points represent the median, while the errorbars represent the 95\% confidence intervals on the distribution of the S/N values. The black dashed line corresponding to the scaling relation in \S\ref{subsec:scaling_rel} is shown for reference.
            For low cross correlation uncertainties, the total and isotropic S/N values have identically large values relative to the anisotropic S/N, implying strong evidence for an isotropic GWB in the data. These S/N values decrease as the cross-correlation uncertainty increases, implying loss of confidence in the detection of a GWB as well as losing the ability to distinguish isoptropy from anisotropy. The anisotropic S/N shows little evolution across uncertainties, though the better fit provided by the additional degrees of freedom in an anisotropic model prevents this S/N from being consistent with zero for small cross-correlation  uncertainties.}
            \label{fig:SN}
        \end{figure}

\begin{figure*}
            \centering
            \includegraphics[width = 1\textwidth]{"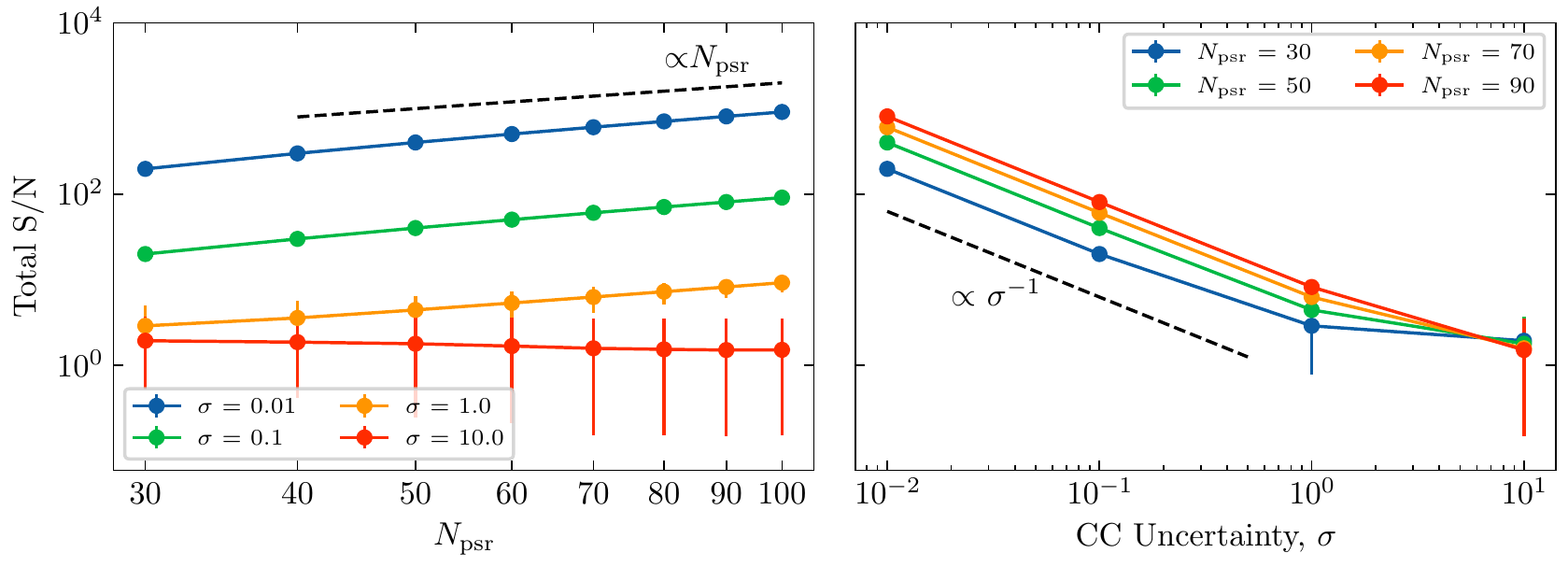"}
            \caption{The evolution of the total S/N values for an ideal PTA with 100 pulsars and an isotropic injected GWB. The points represent the median values across $10^4$ noise realizations, while the errors represent 95\% confidence intervals. 
            \textit{Left:} Evolution of the total S/N as a function of number of pulsars in the PTA for different values of cross-correlation uncertainty. The black dashed line corresponding to the scaling relation in \S\ref{subsec:scaling_rel} is shown for reference.
            \textit{Right:} Evolution of the total S/N as a function of the cross-correlation uncertainty for different number of pulsars in the PTA.
            The black dashed line corresponding to the scaling relation in \S\ref{subsec:scaling_rel} is shown for reference.
            Together, these results show that the total S/N is larger for a PTA with small cross-correlation uncertainties and a large number of pulsars.
            }
            \label{fig:sn_scalings}
        \end{figure*}

        We show that both of these scaling relations for the total S/N are also satisfied when using the non-linear maximum likelihood approach described in \S\ref{subsubsec:sqrt_basis}. 
        \autoref{fig:SN} and the left panel in \autoref{fig:sn_scalings} show that the total S/N scales inversely with the uncertainty on the cross correlations, while the right hand panel of \autoref{fig:sn_scalings} show that the total S/N scales proportionally to the number of pulsars in the PTA. Since the injected signal here is an isotropic GWB, \autoref{fig:SN} shows that the total and isotropic S/N values are identically large, and decrease as the uncertainty on the cross correlations increases. The anisotropic S/N shows little evolution across uncertainties, though the better fit provided by the additional degrees of freedom in an anisotropic model prevent it from being consistent with zero for small  cross-correlation uncertainties. For large uncertainties, \autoref{fig:SN} shows that we lose the ability to detect and distinguish between isotropic or anisotropic signals in the data.
        
        %
        %
        
        \begin{figure}
            \centering
            \includegraphics[width = 0.45\textwidth]{"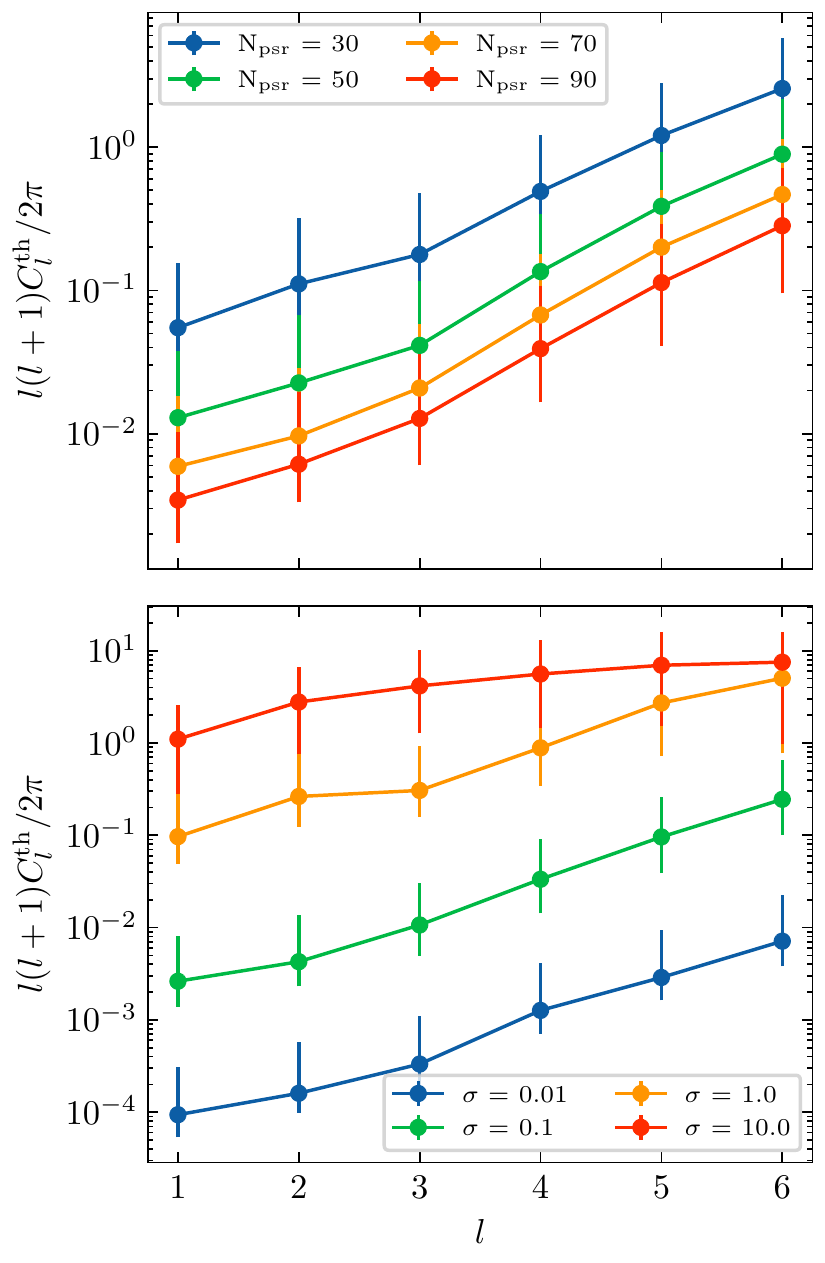"}
            \caption{The evolution of the decision threshold, $C_{l}^{th}$, for an ideal PTA with an isotropic injected GWB. The points represent the median values while the errors represent the 95\% confidence interval values across $10^4$ noise realizations. \textit{Top:} Evolution of the decision threshold per mode for different numbers of pulsars in the PTA. \textit{Bottom:} Evolution of the decision threshold for different cross-correlation uncertainties. These values were generated for an ideal PTA with 100 pulsars and a cross-correlation uncertainty of 0.1. These results show that the decision threshold is lower for PTAs that have a large number of pulsars and smaller cross-correlation uncertainties.}
            \label{fig:dec_thres_scalings}
        \end{figure}
        
        We can similarly compute scaling relations for the decision threshold as a function of the cross correlation uncertainty and number of pulsars in the PTA. Since this is an empirically constructed detection statistic, we do not have analytical expressions for its scaling relations, though we can derive the scaling expressions computationally, as shown in \autoref{fig:dec_thres_scalings}. As expected, as we increase the number of pulsars in the PTA, the decision threshold decreases across all multipoles.  Similarly, as the uncertainty on the cross correlation measurements decreases, so do the multipole-dependent decision thresholds, corresponding to an improved sensitivity to deviations away from isotropy.
    
    \subsection{Sensitivity figure of merit} \label{subsec:summ_stat}
        
        Rather than treating the number of pulsars and the cross-correlation uncertainty as separate variables, we can consider a combination that is inspired by the weighted arithmetic mean of cross-correlation measurements involved in, e.g., S/N calculations. In such calculations, operations like $\sum_{ab}(\cdots)/\sigma_{ab}^2$ are proportional to $N_\mathrm{cc} / \sigma^2 \propto (N_\mathrm{psr}/\sigma)^2$ in the limit of equal cross-correlation measurement uncertainties, or where we can characterise the distribution of uncertainties by its mean or median over pairs. Therefore we define a sensitivity figure of merit (FOM), $N_{\rm psr} / \sigma$, and quantify the dependence of our detection statistics with respect to this. This also allows us to quantify the trade-off between the number of pulsars in the PTA and the cross-correlation uncertainties, which, in turn, are related to the noise characteristics of the pulsars.
        
        \begin{figure}
            \centering
            \includegraphics[width = 0.45\textwidth]{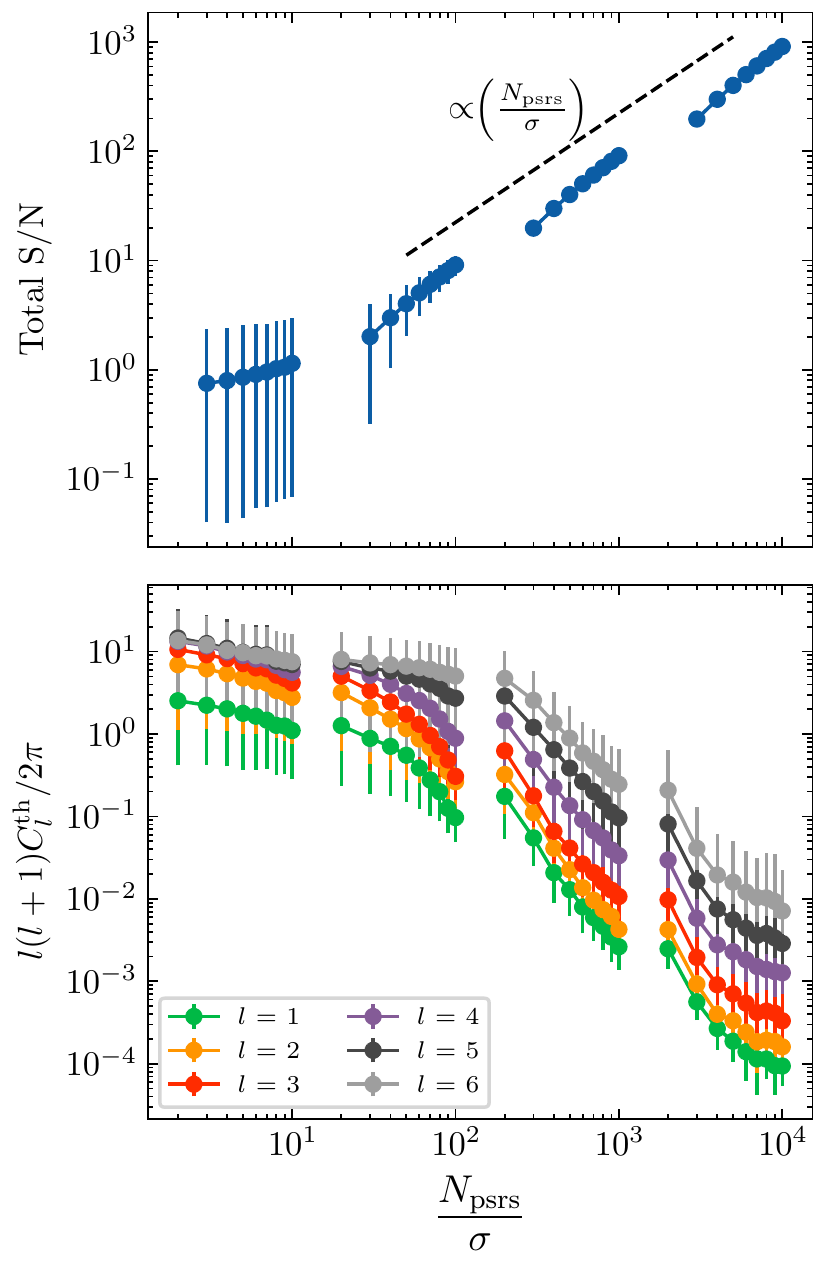}
            \caption{The evolution of detection statistics with the sensitivity FOM, $N_{\rm psrs} / \sigma$, defined in \S\ref{subsec:summ_stat} for an ideal PTA with an isotropic injected GWB. The points represent the medians, while the errorbars represent the 95\% confidence intervals across $10^4$ noise realizations. \textit{Top:} Evolution of the total S/N as a function of the sensitivity FOM. This scaling relation implies that PTAs with either a large number of pulsars or small cross-correlation uncertainties (or both) will return a larger total S/N value than a PTA with fewer pulsars and/or larger cross-correlation uncertainty. \textit{Bottom:} The evolution of the decision threshold as a function of the sensitivity FOM. This implies that PTAs with fewer pulsars or larger cross-correlation uncertainties will have higher decision thresholds, while PTAs with more pulsars and smaller cross-correlation uncertainties will have lower decision thresholds.}
            \label{fig:sum_stat}
        \end{figure}
        
        The relation of the total S/N and decision threshold to the sensitivity FOM is shown in \autoref{fig:sum_stat}. We confirm that the total S/N is proportional to the sensitivity FOM, $N_{\rm psr} / \sigma$ in logarithmic space. This implies that, as expected, fewer pulsars or larger uncertainties on the cross correlation measurements result in a reduced total S/N, while larger numbers of pulsars or smaller uncertainties result in an increase in the total S/N value. Similarly, \autoref{fig:sum_stat} shows the dependence of the decision threshold for each anisotropy multipole on $N_{\rm psr} / \sigma$. Similar to the total S/N, a PTA with fewer pulsars or larger uncertainties on the cross correlation measurements will be able to reject the null hypothesis with lower significance than a PTA with more pulsars and/or smaller uncertainties on the cross correlations. 
    
\section{Simulations with a realistic PTA} \label{sec:sim_real_pta}
    
    The ideal PTA described in \S\ref{sec:sim_ideal_pta} was useful in discerning scaling relationships that map between the PTA design and detection statistics. However, unlike the ideal PTA in \S\ref{sec:sim_ideal_pta}, real PTAs do not (yet) consist of pulsars distributed uniformly across the sky, nor are all the pulsars in the array identical. The latter fact implies that the cross-correlation uncertainties in a real PTA will be described by a distribution, rather than a constant value as assumed in \S\ref{sec:sim_ideal_pta}. 
    
    To simulate a realistic PTA, we use the methods developed in \citet{astro4cast}. We base our simulations on the NANOGrav $12.5$~year dataset \citep{ng12p5_timing}, and extend the dataset to a $20$-year timing baseline to forecast the sensitivity of NANOGrav to anisotropies in the GWB. The TOA timestamps of the initial $12.5$~year portion are the same as those in the real NANOGrav dataset, while the radiometer uncertainties and pulse-phase jitter noise that are injected are obtained from the maximum-likelihood pulsar noise analysis performed as part of the NANOGrav $12.5$~year analysis \citep{NG12p5_detection}. The injection values for the intrinsic per-pulsar red noise were taken from a global PTA analysis that also modeled a common-spectrum process. 
    This is done to isolate the intrinsic red noise in each pulsar's dataset so that it is not contaminated by the common process reported in \citet{NG12p5_detection}.
    
    Once the 45 simulated pulsars from the NANOGrav $12.5$~year dataset are generated using the above recipe, the dataset is then extended into the future by generating distributions for the cadence and measurement uncertainties using the last year's worth of data for each pulsar. We then draw TOAs using these distributions until the dataset has a maximum baseline of $20$~years. Finally, we inject $100$ statistically random realizations of an isotropic gravitational wave background with an amplitude of $A_{\rm GWB} = 2 \times 10^{-15}$ and spectral index $\alpha = -2/3$, consistent with the common process observed in \citet{NG12p5_detection}. 
    
    \begin{figure}
        \centering
        \includegraphics[width = \columnwidth]{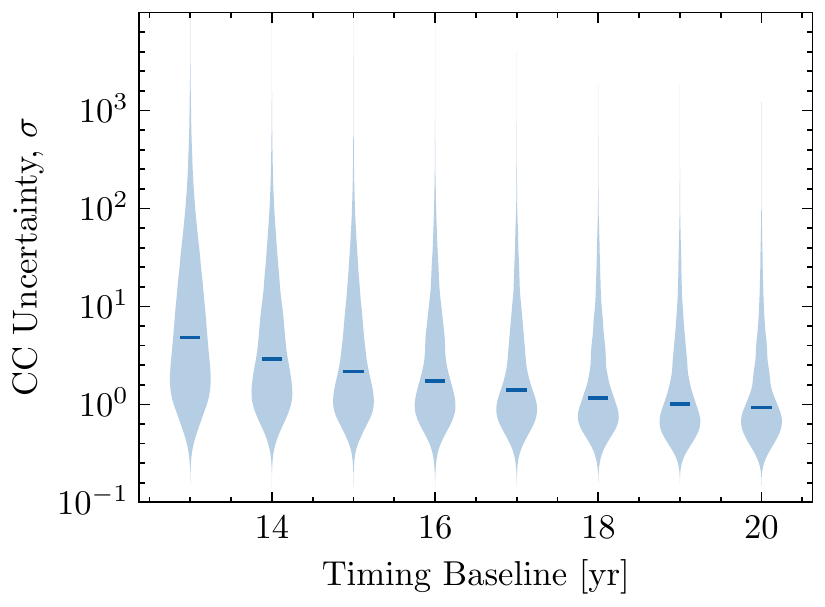}
        \caption{The evolution of the cross-correlation uncertainty across all pulsars and 100 noise realizations of the realistic PTA dataset simulations described in \S\ref{sec:sim_real_pta}. The evolution of the median cross-correlation uncertainty can be approximately described by $\sigma\propto T^{-7/2}$, which is shallower than the scaling law prediction of $\sigma\propto T^{-13/3}$ for the weak-signal regime in \S\ref{subsec:scaling_rel}, but steeper than the strong-signal regime prediction of $\sigma\propto T^{-1/2}$. This implies that the NANOGrav PTA is in the intermediate signal regime, which is corroborated by the fact that the lowest frequencies of the PTA are now dominated by a common-spectrum process (interpreted as the GWB) as shown in \citet{NG12p5_detection}.}
        \label{fig:a4c_sigma}
    \end{figure}
    
    We then pass all 100 realizations of the dataset through the standard NANOGrav detection pipeline \citep{ng11_detection, NG12p5_detection, astro4cast} to calculate the cross correlations and their uncertainties between all pairs in the 45 pulsar dataset (see also \autoref{appendix:real_os}). The evolution of these cross-correlation uncertainties across the 100 realizations as a function of the timing baseline is shown in \autoref{fig:a4c_sigma}. As we can see, the median cross-correlation uncertainty reduces from $\sim$5 at $13$~years (similar to the total baseline of the $12.5$~year dataset) to $\sim$1 at $20$~years, implying a scaling relation $\sigma\propto T^{-7/2}$. This is shallower than the spectral index predicted for the weak-signal regime in \S\ref{subsec:scaling_rel}, which implies that the NANOGrav PTA is in the intermediate-signal regime, which is corroborated by the fact that the lowest frequencies of the PTA are dominated by a common-spectrum process (interpreted as a GWB) as shown in \citet{NG12p5_detection}. 
    Combining the median uncertainty with the 45 pulsars in the PTA, we obtain values for our sensitivity FOM $N_{\rm psr} / \sigma \approx 9$ at $13$~years, and $N_{\rm psr} / \sigma \approx 45$ at $20$~years.
    
    \begin{figure*}
        \centering
        \includegraphics[width = 1\textwidth]{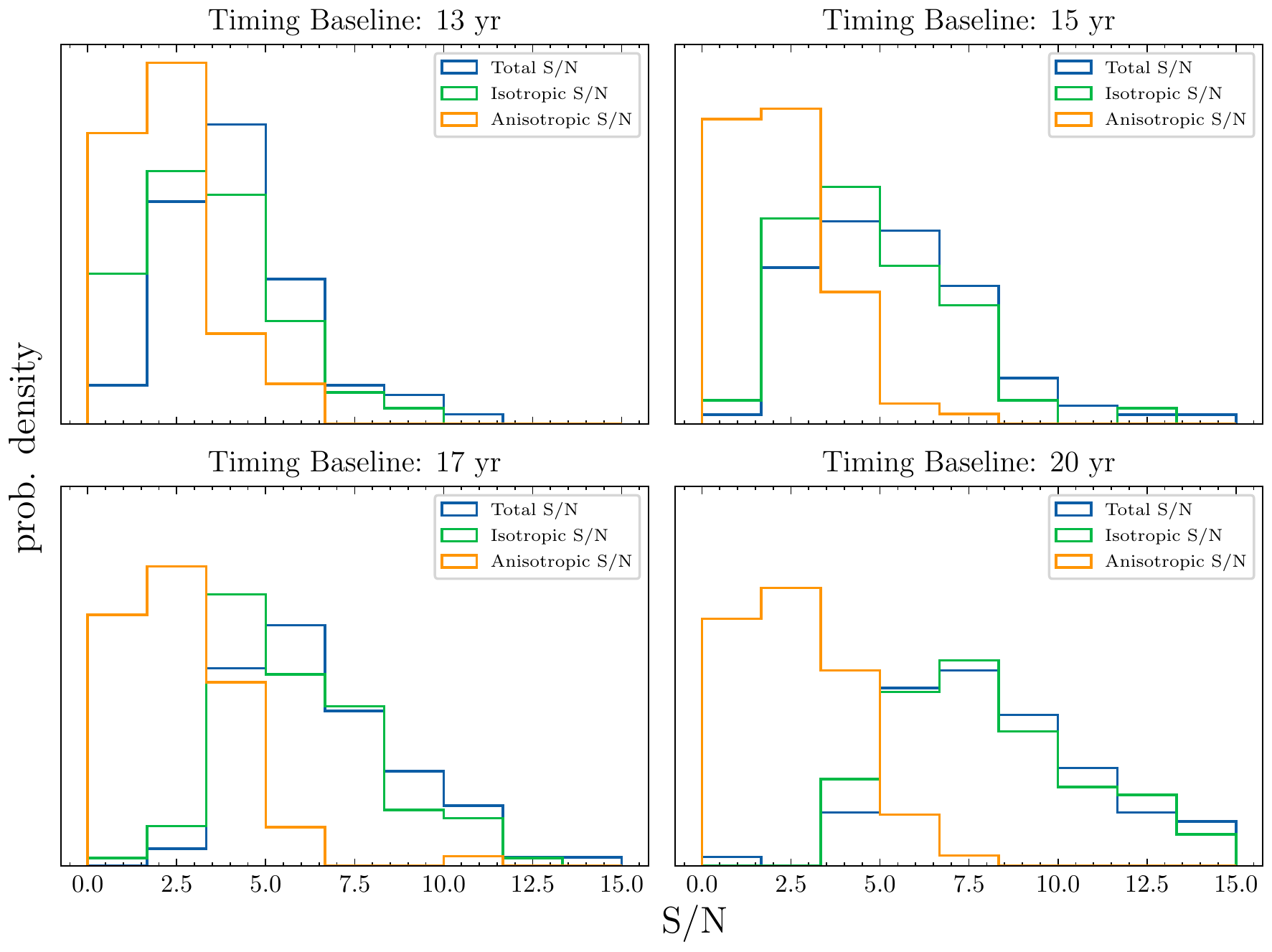}
        \caption{Evolution of the S/N values for the realistic simulations described in \S\ref{sec:sim_real_pta}. The total, isotropic, and anisotropic S/N are shown by the blue, green, and orange histograms, respectively. Since the injected GWB is isotropic, we see the total and isotropic S/N values increase as a function of timing baseline, while the anisotropic S/N stays consistent with zero for all baselines.}
        \label{fig:a4c_sn}
    \end{figure*}
    
    We pass the cross correlations measured from these $100$ realizations through the statistical framework described in \S\ref{sec:methods} to search for the presence of anisotropy under realistic PTA and data-quality conditions. The evolution of the three S/N statistics as a function of time are shown in \autoref{fig:a4c_sn}. Since the injected GWB is isotropic, the total and isotropic S/N increase with the timing baseline. This is consistent with the reduction in the uncertainties on the cross correlations allowing for a stronger detection of the isotropic background. By contrast, the anisotropic S/N does not increase with time, and has support at $\mathrm{S/N}=0$ for all baselines. Note that the total S/N seen in these realistic simulations is consistent with the prediction made in \autoref{fig:sum_stat} and \citet{astro4cast}.
    \begin{figure*}
        \centering
        \includegraphics[width = \textwidth]{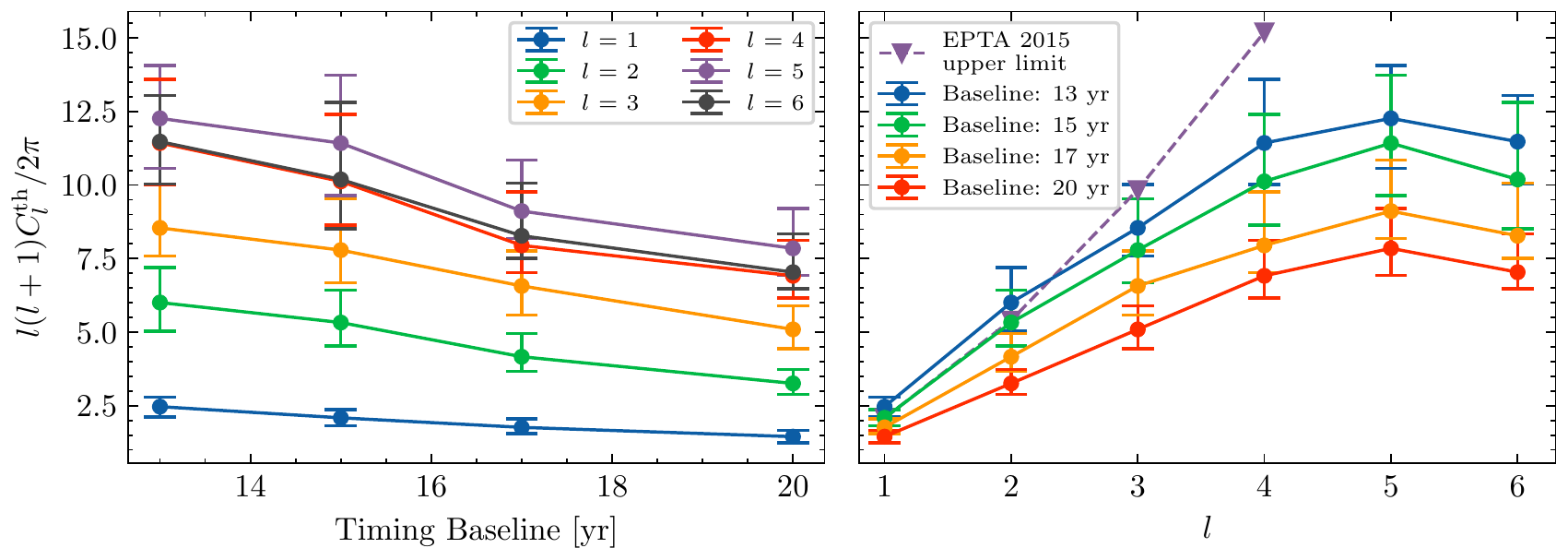}
        \caption{Evolution of decision threshold for realistic simulations described in \S\ref{sec:sim_real_pta}. \textit{Left:} Evolution of the decision threshold as a function of timing baseline for all spherical harmonic multipoles. The decision threshold decreases with an increase in the timing baseline, and higher spherical harmonic multipoles have a higher decision threshold than lower multipoles.
        \textit{Right:} Evolution of the decision threshold across spherical harmonic multipoles for different timing baselines. We also plot the Bayesian 95\% upper limits on anisotropy derived in \citet{epta_anisotropy} from the EPTA Data Release $1$. 
    As these realistic simulations have 45 pulsars with different noise properties resulting in different cross-correlation uncertainties per pulsar pair, we see the sensitivity of the PTA saturate at higher multipoles.}
        \label{fig:a4c_dt}
    \end{figure*}
    Similarly, \autoref{fig:a4c_dt} shows the evolution of the decision threshold as a function of spherical harmonic multipole $l$ and the timing baseline. We find that the anisotropy decision threshold is such that, in terms of the $C_l$ values, GWB anisotropies at levels $C_{l=1} / C_{l=0} \gtrsim 0.3$ (i.e. greater than 30\% of the power in the monopole) would be inconsistent with the null hypothesis of isotropy at the $p = 3\times10^{-3}$ level for the 20 yr baseline. 
    
    For comparison, in \autoref{fig:a4c_dt} we also plot the Bayesian $95\%$ upper limits on GWB anisotropy using six pulsars from the EPTA's first data release \citep{epta_anisotropy} for a model extending to $l_{\rm max} = 4$. This dataset had a maximum baseline of $17.7$~years, which is toward the upper end of the baselines that we simulate for the NANOGrav data. However, the number of pulsars (6) in the EPTA analysis is significantly lower than the number of pulsars in our simulations (45). The longer EPTA timing baseline allows the Bayesian EPTA upper limits and NANOGrav anisotropy decision thresholds to be comparable at low multipoles until NANOGrav's timing baseline exceeds that of the EPTA. However, the larger number of pulsars in NANOGrav not only gives it higher spatial resolution (and thus access to higher multipoles), but also improves the sensitivity of NANOGrav at higher multipoles relative to the EPTA $2015$ limit. As shown in \citet{higher_lmax_limit}, access to these higher multipoles can aid in the localization of GWB anisotropies caused by individual GW sources, or due to finiteness in the source population constituting the GWB. This highlights the importance of including more pulsars in a PTA for spatial resolution, even with the shorter timing baseline of such new additions.
    
\section{Discussion and Conclusion} \label{sec:discuss}
    
    We have explored the detection and characterization of stochastic GWB anisotropy through pulsar cross correlations in a PTA. Using a frequentist maximum likelihood approach, we can search for anisotropy by modeling GWB power in individual sky pixels or through a weighted sum of spherical harmonics. Anisotropy would then manifest in measured cross-correlations between pulsar timing residuals through a power-weighted overlap of pulsar GW antenna response functions. As a refinement on previous approaches, we prevent the GWB power from assuming unphysical negative values by adopting a model that naturally restricts this; we have referred to this as the \textit{square-root spherical harmonic basis} throughout our analysis. We have also defined two detection metrics: $(a)$ the signal-to-noise ratio, S/N, defined as the ratio between the maximum likelihood values between a signal and noise model, and $(b)$ the anisotropy decision threshold, $C_{l}^\mathrm{th}$, defined as the level at which the measured angular power is inconsistent with isotropy at the $p = 3\times10^{-3}$ ($\sim3\sigma$) level. The S/N comes in three flavors: $(i)$ the total S/N, which measures the strength of an anisotropic GWB model against noise alone; $(ii)$ the isotropic S/N which measures the strength of an isotropic GWB model against noise alone (this directly corresponds to the usual optimal statistic S/N used in PTA data analysis); and $(iii)$ the anisotropic S/N, which measures the statistical preference for anisotropy against isotropy.
    
    We examined the evolution of these detection statistics as a function of the uncertainty on the measured cross correlations, as well as on the number of pulsars in the PTA. We showed that the cross-correlation uncertainty and the number of pulsars in a PTA can be combined into a single figure of merit for the PTA sensitivity, $N_{\rm psrs} / \sigma$, which succintly maps the PTA configuration and noise specifications to the detectability of anisotropy. Our scaling relations show that increasing the number of pulsars in an array, while reducing the uncertainty on the cross correlation measurements, leads to higher total S/N and lower anisotropy decision thresholds. As shown in \S\ref{sec:connection}, the cross-correlation uncertainty scales inversely with both the timing baseline and cadence of observation for each pulsar, with the power-law index dependent on the signal regime occupied by the PTA. The pulsar timing baseline is set to increase as PTAs continue operation into the future, as well as through IPTA data combinations \citep[e.g.,][]{ipta_dr2_dataset}. Improving observing cadence is more challenging due to constraints on the available telescope time for each PTA, though once again IPTA data combinations can help alleviate this problem. In addition to IPTA data combinations, the CHIME radio telescope \citep{CHIME_pulsar} will offer $\sim$daily cadence to the NANOGrav PTA \citep{NANOGrav}, which is an order of magnitude improvement over the current $\sim$monthly cadence employed by NANOGrav.
    
    Finally, we examined the evolution of the S/N and anisotropy decision thresholds as a function of timing baseline using realistic NANOGrav data. Since we injected an isotropic GWB signal in these data, we found that the anisotropic S/N remains consistent with zero at all times and across all signal realizations. By contrast, the total and isotropic S/N increase with time, as expected \citep{siemens_scaling_laws}. We find that any anisotropic GWB power distribution with $C_{l = 1} \gtrsim 0.3 C_{l = 0}$ would be in tension with an isotropic model at the $p = 3\times10^{-3}$ ($\sim3\sigma$) level. We note that these simulations held the number of pulsars in the array fixed to the $45$ that were included in the NANOGrav $12.5$~year dataset. However, this number will increase in future NANOGrav datasets. Based on results in \S\ref{subsec:summ_stat}, this will lead to larger total S/N values and lower anisotropy decision thresholds, implying improved sensitivity to any anisotropy that might be present in the real GWB. IPTA data combinations will further increase the timing baseline and number of pulsars in the array allowing further improvements on the ability to detect anisotropy. Furthermore, new instruments such as ultra-wideband receivers, and new telescopes such as MeerKAT \citep{meerkat_psr}, SKA \citep{SKAPTA}, and DSA-2000 \citep{dsa2000} will also aid in the detection and characterization of anisotropy with PTAs.
    
    The techniques and scaling relations that we have developed in this work are PTA-agnostic and can be projected onto any PTA specification, allowing for immediate usage by the broader PTA community. Yet we have made assumptions in our framework that can be generalized in the future. For example, while our techniques operate on PTA data at the level of cross correlations rather than TOAs, in order to get to that stage we have implicitly assumed that the GWB characteristic strain spectrum is well-described by a power-law model. This follows the same approach as \citet{NG12p5_detection}, where an average power-law spectrum $h_c\propto f^{-2/3}$ is assumed for the GWB. For a GWB produced by a population of inspiraling SMBHBs, this power-law representation is an approximation to the true spectrum \citep{phinney}, where there are different SMBHBs contributing to the GWB at different frequencies \citep{Kelley_real_SMBHB_spectrum}. Thus, a more appropriate way to characterize anisotropy in the SMBHB GW background would be to measure the cross correlations of pulsar TOAs as a function of GW frequency, rather than our current approach of computing cross correlations that are filtered against a power-law GWB spectral template (see \autoref{appendix:real_os}). We would then have a more general data structure that includes pulsar cross-correlations and uncertainties for each GW frequency, for which the methods developed here can be applied at each of those frequencies independently. We also note that the methods developed here can be modified to search for multiple backgrounds \citep{multiple_anis_bkgnd}, where an astrophysical background \citep[e.g. from SMBHBs,][]{Kelley_real_SMBHB_spectrum, Sesana2004, BurkeSpolaor2019} would be expected to be anisotropic, while a cosmological background \citep[e.g. from cosmic strings,][]{olmez_anisotropy} may be isotropic. 
    
    We plan to explore these improvements and generalizations in future analyses in a bid to extract as much spatial and angular information as possible from the exciting new PTA datasets now under development. As mentioned, these techniques will not only aid in the detection of GWB anisotropy, but also in its characterization for the purposes of isolating regions of excess power that may be indicative of individually-resolvable GW sources, and as leverage for the separation of potentially multiple stochastic GW signals of astrophysical and cosmological origin.


\begin{acknowledgements}
    
    NP was supported by the Vanderbilt Initiative in Data Intensive Astrophysics (VIDA) Fellowship. SRT acknowledges support from NSF AST-200793, PHY-2020265, PHY-2146016, and a Vanderbilt University College of Arts \& Science Dean's Faculty Fellowship. JDR was partially supported by start-up funds provided by Texas Tech University.
    
    This work has been carried out by the NANOGrav collaboration, which is part of the International Pulsar Timing Array. 
    The NANOGrav project receives support from National Science Foundation (NSF) Physics Frontiers Center award number \#1430284 and \#2020265.
    
    This work was conducted using the resources of the Advanced Computing Center for Research and Education at Vanderbilt University, Nashville, TN.
    
\section*{Software:} 
We use \texttt{lmfit} \citep{lmfit} for the non-linear least-squares minimization in the square-root spherical harmonic basis. 
We use \texttt{libstempo} \citep{libstempo} to generate our realistic PTA datasets and to inject the pulsar noise parameters and GWB signals in these datasets. 
We use the software packages \texttt{Enterprise} \citep{enterprise} and \texttt{enterprise\_extensions} \citep{enterpriseextensions} for model construction, along with \texttt{PTMCMCSampler} \citep{PTMCMC} as the Markov Chain Monte Carlo sampler for our realistic PTA Bayesian analyses. 
We also extensively used Matplotlib \citep{Matplotlib2007}, NumPy \citep{Numpy2020}, Python \citep{Python2007,Python2011}, and SciPy \citep{Scipy2020}.
    
\end{acknowledgements}

\appendix

\section{Computing cross-correlations in realistic pulsar-timing datasets} \label{appendix:real_os}

Reiterating \autoref{eq:cross_corr}, the cross-correlation value (and its uncertainty) measured between pulsar-$a$ and pulsar-$b$ are
\begin{align}
    \displaystyle \rho_{ab} &= \frac{\delta\textbf{t}^T_{a} \textbf{P}^{-1}_{a} \hat{\textbf{S}}_{ab} \textbf{P}^{-1}_{b} \delta\textbf{t}^T_{b}}{\textrm{tr}\left[ \textbf{P}^{-1}_{a} \hat{\textbf{S}}_{ab} \textbf{P}^{-1}_{b} \hat{\textbf{S}}_{ba} \right]}, \nonumber\\
    \displaystyle \sigma_{ab} &= \left( \textrm{tr}\left[ \textbf{P}^{-1}_{a} \hat{\textbf{S}}_{ab} \textbf{P}^{-1}_{b} \hat{\textbf{S}}_{ba} \right] \right)^{-1/2},
\end{align}
where $\delta\mathbf{t}_a$ is a vector of timing residuals for pulsar-$a$, $\mathbf{P}_a = \langle \delta \mathbf{t}_a \delta \mathbf{t}_a^T\rangle$ is the measured autocovariance matrix of pulsar-$a$, and $\hat{\textbf{S}}_{ab}$ is the template scaled-covariance matrix between pulsar-$a$ and pulsar-$b$. This scaled-covariance matrix is a template for the spectral shape only, and is amplitude and ORF independent. It is related to the full covariance matrix by $\textbf{S}_{ab} = \langle \delta \mathbf{t}_a \delta \mathbf{t}_b^T\rangle = A^2\chi_{ab}\hat{\textbf{S}}_{ab}$.

Covariance matrices in the PTA likelihood are modeled using rank-reduced approximations, e.g., long-timescale stochastic processes are modeled on a truncated Fourier basis with a number of frequencies that is much smaller than the number of TOAs. The induced timing delays of such stochastic processes are modeled as $\mathbf{r}=\mathbf{T}\mathbf{b}$, where $\mathbf{b}$ is a vector of amplitude coefficients for the basis design matrix $\mathbf{T}=[\mathbf{M}\quad\mathbf{F}]$, which itself is a concatenation of the stablized timing-model design matrix $\mathbf{M}$ and Fourier design matrix $\mathbf{F}$. The former is a matrix of partial derivatives of TOAs with respect to timing model parameters, which is then column-normed or SVD-stablized to reduce the dynamic range of the entries. The latter is a matrix of sine and cosine basis functions evaluated at all TOAs for each sampling frequency of the time series. The autocovariance matrix of pulsars is then modeled as $\mathbf{P}_a = \mathbf{N}_a + \mathbf{T}_a\mathbf{B}_{aa}\mathbf{T}_a^T$, where $\mathbf{N}_a$ is the white-noise covariance matrix of pulsar-$a$, and
\begin{equation}
    \vb{B} = \mqty[\boldsymbol{\infty} & \vb{0} \\ \vb{0} & \boldsymbol{\phi} ] 
\end{equation}
is the covariance matrix of the $\mathbf{b}$ coefficients, $\mathbf{B} = \langle \mathbf{b}\mathbf{b}^T\rangle$, with infinite variance in the timing model diagonal entries to approximate an unbounded uniform prior, and $\boldsymbol{\phi}$ as the variance of the Fourier coefficients that is related to the power spectral density of the stochastic process. The inter-pulsar covariance matrix is treated similarly, except that the timing model entries can be completely ignored since they will be uncorrelated between different pulsars. Hence, $\hat{\textbf{S}}_{ab} = \mathbf{F}_a \boldsymbol{\hat{\phi}} \mathbf{F}_b^T$, where $\boldsymbol{\hat{\phi}}$ is the scaled Fourier covariance matrix of the GWB (i.e., amplitude and ORF independent).

We use the Woodbury matrix lemma to perform inversions, such that
\begin{align}
    \mathbf{P}_a^{-1} &= \left( \mathbf{N}_a + \mathbf{T}_a\mathbf{B}_{aa}\mathbf{T}_a^T \right)^{-1} \nonumber\\
    &= \mathbf{N}_a^{-1} - \mathbf{N}_a^{-1}\mathbf{T}_a\left( \mathbf{B}_{aa}^{-1} + \mathbf{T}_a^T\mathbf{N}_a^{-1}\mathbf{T}_a \right)^{-1}\mathbf{T}_a^T\mathbf{N}_a^{-1} \nonumber\\
    &= \mathbf{N}_a^{-1} - \mathbf{N}_a^{-1}\mathbf{T}_a\boldsymbol{\Sigma}_a^{-1}\mathbf{T}_a^T\mathbf{N}_a^{-1},
\end{align}
%
%
Hence the numerator of \autoref{eq:cross_corr} can be written as
\begin{equation}
    \delta\textbf{t}^T_{a} \textbf{P}^{-1}_{a} \hat{\textbf{S}}_{ab} \textbf{P}^{-1}_{b} \delta\textbf{t}^T_{b} =  \mathbf{X}_a^T \hat{\boldsymbol{\phi}}  \mathbf{X}_b,
\end{equation}
where
\begin{equation}
    \mathbf{X}_a = \mathbf{F}_a^T\mathbf{N}_a^{-1}\delta\textbf{t}_a  - \mathbf{F}_a^T\mathbf{N}_a^{-1}\mathbf{T}_a\boldsymbol{\Sigma}_a^{-1}\mathbf{T}_a^T\mathbf{N}_a^{-1}\delta\textbf{t}_a,
\end{equation}
and the denominator is written as
\begin{equation}
    \textrm{tr}\left[ \textbf{P}^{-1}_{a} \hat{\textbf{S}}_{ab} \textbf{P}^{-1}_{b} \hat{\textbf{S}}_{ba} \right] = \textrm{tr}\left[ \mathbf{Z}_a \hat{\boldsymbol{\phi}} \mathbf{Z}_b \hat{\boldsymbol{\phi}}\right],
\end{equation}
where
\begin{equation}
    \mathbf{Z}_a = \mathbf{F}_a^T \mathbf{N}_a^{-1} \mathbf{F}_a - \mathbf{F}_a^T \mathbf{N}_a^{-1} \mathbf{T}_a\boldsymbol{\Sigma}_a^{-1}\mathbf{T}_a^T\mathbf{N}_a^{-1} \mathbf{F}_a.
\end{equation}

Crucially, the $\mathbf{X}$ and $\mathbf{Z}$ matrices for each pulsar depend on measured noise characteristics (e.g., white noise and intrinsic red noise) or fixed aspects of modeling (e.g., timing-model and Fourier design matrices). They can thus be constructed and cached for follow-up analysis. The diagonal matrix $\hat{\boldsymbol{\phi}}$ acts as a spectral template that we use in a procedure akin to matched filtering, where we assess the S/N of cross-correlated signals with, e.g., a power-law PSD that has a $-13/3$ exponent, matching expectations for a circular, GW-driven population of SMBHBs forming a stochastic GWB.

Thus, in production-level PTA searches for the stochastic GWB, the cross-correlations are computed as
\begin{align}
    \displaystyle \rho_{ab} &= \frac{\mathbf{X}_a^T \hat{\boldsymbol{\phi}} \mathbf{X}_b}{\textrm{tr}\left[ \mathbf{Z}_a \hat{\boldsymbol{\phi}} \mathbf{Z}_b \hat{\boldsymbol{\phi}}\right]}, \nonumber\\
    \displaystyle \sigma_{ab} &= \left( \textrm{tr}\left[ \mathbf{Z}_a \hat{\boldsymbol{\phi}} \mathbf{Z}_b \hat{\boldsymbol{\phi}} \right] \right)^{-1/2}.
\end{align}

\section{Cross-correlation measurement uncertainty from PTA specifications} \label{appendix:unc_scaling}
    
    As shown in \S\ref{sec:connection}, the trace in \autoref{eq:cross_corr} can be written as
    \begin{equation}
        \displaystyle \textrm{tr}\left[ \textbf{P}^{-1}_{a} \hat{\textbf{S}}_{ab} \textbf{P}^{-1}_{b} \hat{\textbf{S}}_{ba} \right] = \frac{2 T}{A^4} \int_{f_l}^{f_h} df \frac{P_g^2(f)}{P_a(f) P_b(f)},
        \label{eq:trace_adx}
    \end{equation}
    where $T$ is the timing baseline; $A$ is the amplitude of the GWB; $P_g(f)$ is the power spectral density of timing residuals induced by the GWB; $P_a(f)$ and $P_b(f)$ are the intrinsic power spectral density of timing residuals in pulsars $a$ and $b$; and $f_l = 1/T$ and $f_h$ are the low and high frequency cutoffs used in the GWB analysis. For a GWB with characteristic strain spectrum 
    \begin{equation}
        \displaystyle h_c(f) = A \left( \frac{f}{f_{\rm yr}} \right)^{\alpha},
        \label{eq:char_strain}
    \end{equation}
    and using the convention from \citet{siemens_scaling_laws}, the GWB spectrum can be written in the timing residual space as
    \begin{equation}
        \displaystyle P_g(f) = \frac{A^2}{12 \pi^2} \left( \frac{f}{f_{\rm ref}} \right)^{2\alpha} f^{-3} = b f^{-\gamma},
        \label{eq:gwb_spec_adx}
    \end{equation}
    where $\gamma = 3 - 2\alpha$, and $f_{\rm ref}$ is a reference frequency. Consequently, the uncertainty on the cross correlations can be written as
    \begin{equation}
        \displaystyle \sigma_{ab} = \frac{A^2}{\sqrt{2 T}} \left[ \int_{f_l}^{f_h} df \frac{P_g^2(f)}{P_a(f) P_b(f)} \right]^{-1/2}.
        \label{eq:base_cc_unc}
    \end{equation}
    %
    
    As described in \citet{siemens_scaling_laws}, assuming no intrinsic red noise in the pulsars, and all pulsars are identical, the intrinsic power can be written as the sum of the GWB power and the white noise, $w$, in each pulsar, $P_a(f) = P_b(f) = P_g(f) + 2 w^2 \Delta t$, where the cadence of pulsar observations is given by $c = 1 / \Delta t$. Substituting the intrinsic and GWB power in \autoref{eq:base_cc_unc}, the cross correlation uncertainty can be written as
    \begin{equation}
        \displaystyle \sigma_{ab} = \frac{A^2}{\sqrt{2 T}} \left[ \int_{f_l}^{f_h} df \frac{b^2 f^{-2 \gamma}}{(bf^{-\gamma} + 2 w^2 \Delta t)^2} \right]^{-1/2}.
        \label{eq:explicit_base_cc_unc}
    \end{equation}
    
    The solution for the integral in \autoref{eq:explicit_base_cc_unc} is non-trivial, but as described in \citet{siemens_scaling_laws} and \citet{optimal_statistic_chamberlin}, it can be solved using hypergeometric functions. We can define three regimes in which PTAs can operate: (i) the weak-signal regime, where the intrinsic pulsar noise dominates over the GWB power, i.e. $2 w^2 \Delta t \gg bf^{-\gamma}$; (ii) the strong-signal regime, where the GWB power dominates the intrinsic white noise, i.e. $2 w^2 \Delta t \ll bf^{-\gamma}$; (iii) the intermediate-signal regime where only the lowest few frequency bins in the PTA are dominated by the GWB, while at higher frequencies, the white noise dominates the GWB power. Given the recent results from the regional PTAs \citep{NG12p5_detection, epta_dr2_gwb, ppta_dr2_gwb} and the IPTA \citep{ipta_dr2_gwb}, currently PTAs are in the intermediate-signal regime.
    
    In the weak-signal regime, \citet{siemens_scaling_laws} showed that \autoref{eq:explicit_base_cc_unc} can be written as
    \begin{align}
        \displaystyle \sigma_{ab} &\approx \frac{A^2}{\sqrt{2 T}} \left[ \frac{b^2}{(2 w^2 \Delta t)^2} \frac{T^{2\gamma - 1}}{2\gamma - 1} \right]^{-1/2}
    \end{align}
    which reduces to
    \begin{align}
        \sigma_{ab} &\approx 12 \pi^2 \frac{w^2}{c}\sqrt{4\gamma-2}\times T^{-\gamma}.
        \label{eq:weak_sig_unc}
    \end{align}

    Similarly, in the strong-signal regime, the same integral reduces to \citep{siemens_scaling_laws}
    \begin{equation}
        \displaystyle \sigma_{ab} \approx \frac{1}{2} \frac{A^2}{\sqrt{c T}}.
        \label{eq:strong_sig_unc}
    \end{equation}
    
    Note that the scaling expressions above include the GWB amplitude, while in the rest of this paper, we use amplitude-scaled cross correlation values and uncertainties.

\bibliography{bib.bib}
\bibliographystyle{aastex}

\end{document}